# Using Uncertainty in Deep Learning Reconstruction for Cone-Beam CT of the Brain

P. Wu,[1] A. Sisniega,[1] A. Uneri,[1] R. Han,[1] C. K. Jones,[1] P. Vagdragi,[1] X. Zhang,[1]
M. Luciano,[2] W. S. Anderson,[2] and J. H. Siewerdsen[1,2]

**1** Department of Biomedical Engineering, Johns Hopkins University, Baltimore MD USA 21205
**2** Department of Neurosurgery, Johns Hopkins Medical Institute, Baltimore MD USA 21287

**Purpose:** Contrast resolution beyond the limits of conventional cone-beam CT (CBCT) systems is essential to high-quality imaging of the brain for image-guided neurosurgery. We present a deep learning reconstruction method (dubbed DL-Recon) that integrates physically principled reconstruction models with DL-based image synthesis based on the statistical uncertainty in the synthesis image.

**Methods:** A synthesis network was developed to generate a synthesized CBCT image (DL-Synthesis) from an uncorrected filtered back-projection (FBP) image. To improve generalizability (including accurate representation of lesions not seen in training), voxel-wise epistemic uncertainty of DL-Synthesis was computed using a Bayesian inference technique (Monte-Carlo dropout). In regions of high uncertainty, the DL-Recon method incorporates information from a physics-based reconstruction model and artifact-corrected projection data. Two forms of the DL-Recon method are proposed: (i) image-domain fusion of DL-Synthesis and FBP (denoted DL-FBP) weighted by DL uncertainty; and (ii) a model-based iterative image reconstruction (MBIR) optimization using DL-Synthesis to compute a spatially varying regularization term based on DL uncertainty (denoted DL-MBIR). A high-fidelity forward simulator was developed to provide physically realistic simulated CBCT images over a broad range of exposure conditions as training and testing data for the synthesis network. The performance of DL-Recon was investigated using CBCT images with simulated and real low-contrast lesions in the brain.

**Results:** The error in DL-Synthesis images was correlated with the uncertainty in the synthesis estimate. Compared to FBP and PWLS, the DL-Recon methods (both DL-FBP and DL-MBIR) showed ~50% reduction in noise (at matched spatial resolution) and ~40-70% improvement in image uniformity. Conventional DL-Synthesis alone exhibited ~10-60% under-estimation of lesion contrast and ~5-40% reduction in lesion segmentation accuracy (Dice coefficient) in simulated and real brain lesions, suggesting a lack of reliability / generalizability for structures unseen in the training data. DL-FBP and DL-MBIR improved the accuracy of reconstruction by directly incorporating information from the measurements in regions of high uncertainty. Both maintained the advantages of DL-Synthesis. DL-FBP offered the runtime efficiency of FBP, and DL-MBIR offered a further ~10% improvement in contrast resolution compared to DL-FBP.

**Conclusion:** The image quality and robustness of CBCT of the brain were greatly improved with the proposed DL-Recon method incorporating uncertainty estimation with physically principled reconstruction models. Translation to clinical studies is underway.

## 1 Introduction

Cone-beam CT (CBCT) is increasingly prevalent in image-guided neurosurgery. Many implementations, however, are only suitable to visualization of high-contrast bone or surgical instrumentation. Challenges to imaging of low-contrast soft tissues are well established, including artifacts (e.g., x-ray scatter, beam-hardening)[1,2] and quantum and electronic noise that further limit contrast resolution.[2]

Recent advances in deep learning (DL) based reconstruction have opened the possibility for improved contrast resolution in CBCT.[3,4] A popular approach to DL-based reconstruction involves generation of a post-processed image from input given by conventional reconstruction [e.g., filtered backprojection (FBP) or model-based iterative reconstruction (MBIR)].[3,4] While DL methods provide a powerful tool for image synthesis, their accuracy and generalizability may not be guaranteed. Inaccuracy can arise especially when the input deviates strongly from the training cohort (e.g., pathology or imaging conditions not included in the training dataset). This is especially true for image-domain post-processing methods[3,4] that do not explicitly enforce fidelity to the projection data.

Important gains in the performance of DL reconstruction can be achieved by means of a principled approach that invokes understanding of mechanistic physical models underlying the data and/or the reconstruction method. Some researchers incorporate physical models by using the DL synthesized image as a prior (regularization) term in MBIR.[5,6] Such an approach permits deviations from the prior, as enforced by the data fidelity term, although conventional spatially invariant weighting of the regularization could underweight contributions of the prior in some regions and overweight the prior in regions in which the prior deviates from the image data due to inaccuracies in the DL synthesis image.

In this work, a DL reconstruction method (denoted DL-Recon) is presented. The method integrates physical models with image synthesis in a spatially varying manner. A Bayesian inference technique[7] is used to compute voxel-wise uncertainty in the DL synthesis image. In regions where the uncertainty is high, the method leverages more contribution of the measured data, using either (i) FBP reconstruction (denoted DL-FBP); or (ii) a physics-based optimization model as in MBIR (denoted DL-MBIR). Thus, the contributions of both DL and physics model-based methods are leveraged in a physically principled manner for improved overall performance and reliability. The performance of the DL-Recon methods was validated in studies with CBCT images involving simulated and real low-contrast lesions in the brain.

## 2 Materials and Methods

As illustrated in Fig. 1, the proposed DL-Recon method involves three steps: (i) With an uncorrected FBP image ($\mu_{init}$) as input, the synthesis network generates a synthetic CBCT image (denoted DL-Synthesis, $\mu_s$) _and_ an uncertainty map ($\sigma$); (ii) Input projections $y$ are corrected for artifacts. In this work, $\mu_s$ is taken as the object model for correction of x-ray scatter and beam-hardening effects;[1,2] and (iii) Information from the corrected projection data and a physics-based reconstruction model (FBP or MBIR) is integrated with $\mu_s$ in relation to the uncertainty map ($\sigma$) to yield the final reconstruction output, $\mu$ (DL-Recon).

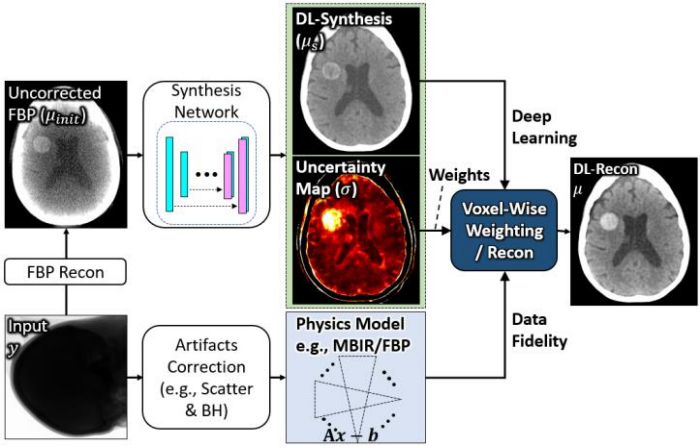

Figure 1. Illustration of the proposed DL-Recon method incorporating DL-Synthesis and uncertainty information with a physics-based model. The DL-Recon result combines the performance of DL image synthesis with the reliability of physics-based models (FBP or MBIR).

## 2.1 Uncertainty Estimation in DL Image Synthesis

Following the work of Gal et al.,[7] the variance of the network output is taken as a proxy for predictive uncertainty. Predictive uncertainty is interpreted in two forms that separately describe epistemic uncertainty due to noise in the network parameters (weights) and aleatoric uncertainty due to noise in the training data. The work reported below focuses on epistemic uncertainty which is associated with a lack of information available in the training data (e.g., previously unseen pathology).

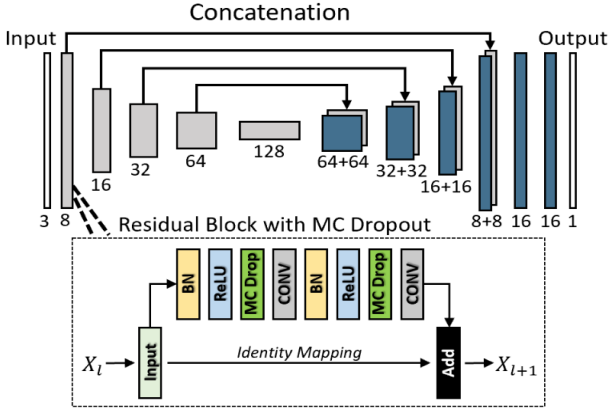

Figure 2. CNN architecture for image-domain synthesis. Epistemic uncertainty is estimated via Monte Carlo (MC) dropout layers (dropout rate = 0.2) inserted in the downsampling and upsampling branch.

The synthesis network used in this work is a residual U-Net[8] type of network as shown in Fig. 2. During network training, dropout is used to fit an approximate distribution over the weights of the CNN (Bayesian inference).[7] Then, during inference, dropout is applied in a MC manner to draw samples of the weights from the fitted approximate distribution. Inference is then performed multiple times, each with a different weights sample from the MC dropout. The variance of the network output (epistemic uncertainty) is then estimated from the sample variance:

$$\sigma^2 = \text{Var}(\mu_s) \approx \frac{1}{N} \sum_{n=1}^{N} f(\mu_{init}, P_n) f(\mu_{init}, P_n)^T -$$
$$\left( \frac{1}{N} \sum_{n=1}^{N} f(\mu_{init}, P_n) \right) \left( \frac{1}{N} \sum_{n=1}^{N} f(\mu_{init}, P_n) \right)^T \quad (1)$$

where $N$ is the number of weights samples (i.e., number of inferences, 16 in this work), and $\mu_s = f(\mu_{init}, P_n)$ is the network output for input $\mu_{init}$ and weights sample $P_n$.

## 2.2 The DL-Recon Methods

### 2.2.1 DL-Recon Method #1: DL-FBP

First, a straightforward approach is presented to integrate physics-based information with DL-Synthesis as a function of the (spatially varying) uncertainty via weighted fusion with an FBP reconstruction in the image domain:

$$\mu = \beta \odot \mu_s + (1 - \beta) \odot \mu_{fbp} \quad (2)$$
$$\beta = \left( \frac{[\sigma_m - D(\sigma)]_+}{\sigma_m} \right)^p \quad (3)$$

where $\mu_{fbp}$ is an (artifact-corrected) FBP reconstruction, and $\odot$ represents voxel-wise multiplication. The relative contribution of $\mu_s$ and $\mu_{fbp}$ to each voxel is controlled by the spatially varying map $\beta$ [Eq. (3)] ranging 0 to 1. The $\beta$ map is shaped by the scalar exponent $p$ [with $p = 2$ in this work]. The uncertainty ($\sigma$) of the synthesis ($\mu_s$) thus yields a normalized $\beta$ map, where $\sigma_m$ is the maximum allowed uncertainty (i.e., $\beta = 0$ for $\sigma > \sigma_m$). A dilation operator $D$ (5-voxel dilation in this work) was used to promote over-estimation of the uncertain region and ensure smooth-transition of the $\beta$ map. In this way, contributions from the physics-based / analytical reconstruction image ($\mu_{fbp}$) are greater where DL-Synthesis is less reliable (high uncertainty).

### 2.2.2 DL-Recon Method #2: DL-MBIR

Second, we propose integration of the DL synthesis result with a physically principled model via MBIR[9] — for example, iterative optimization of an objective combining the uncertainty-weighted DL-Synthesis based prior with a penalized weighted-least squares (PWLS)[9] estimate:

$$\hat{\mu} = \arg\min_{\mu} \frac{1}{2} \|A\mu - l\|_W^2 + \lambda \ \|\Psi(\mu)\|_1 + \lambda_{DL}\Psi_{DL} \quad (4)$$
$$\Psi_{DL} = \beta \odot \|\mu - \mu_s\|_1 \quad (5)$$

The first two terms in Eq. (4) are recognized simply as PWLS with an image roughness penalty (quadratic penalty in this work), where the system matrix $A$ denotes the linear forward projection operator, $l$ is the (artifact-corrected) line integral, W is the estimated variance for each measurement, and $\|\Psi(\mu)\|_1$ is the roughness penalty based on neighborhood differences with a scalar weighting $\lambda$ . The third term (i.e., the "deep learning" term) serves as an additional penalty on differences between the DL-Synthesis and the current estimate ($\mu$). The global contribution of this penalty is controlled by a constant scalar $\lambda_{DL}$, and – importantly – the spatially varying penalty strength is controlled by the uncertainty map via $\beta$ [Eq. (3)]. Thus, the penalty strength is inversely proportional to the uncertainty of DL-Synthesis, allowing greater contribution from the physics-based data fidelity term where DL-Synthesis is less certain. Compared to DL-FBP, DL-MBIR allows incorporation of an explicit projection domain data fidelity constraint and more accurate physics models.

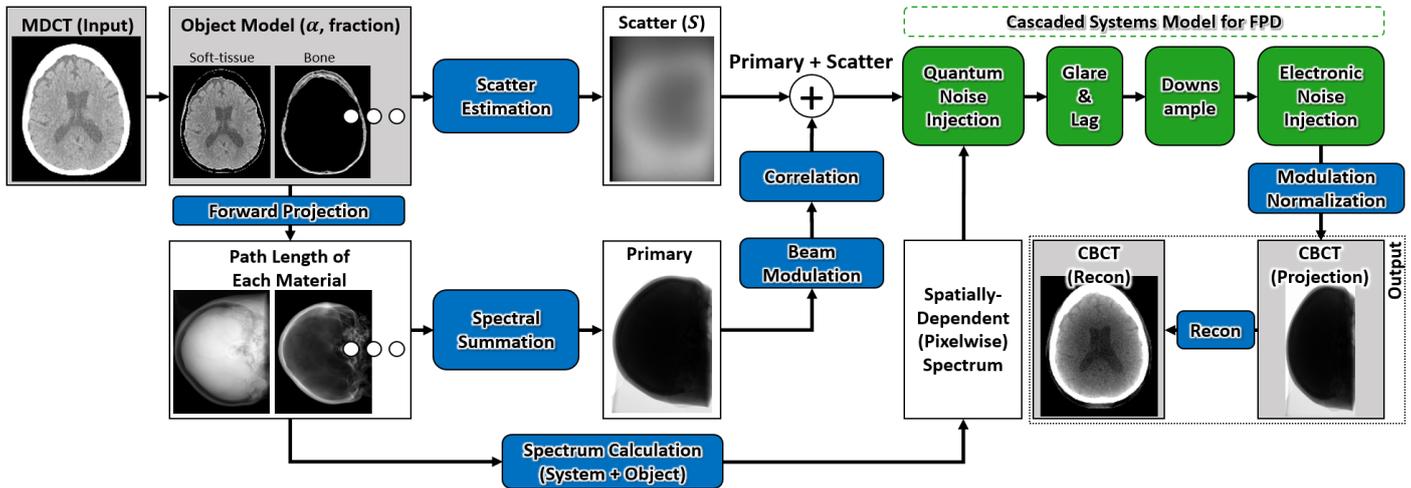

*Figure 3. Flowchart of the high-fidelity forward simulator. Input to the simulator is the high quality MDCT volumes. Output is the simulated CBCT.*

## 2.3 Training Data Generation

A high-fidelity forward simulator (Fig. 3) was developed to simulate realistic CBCT projection data from high-quality, helical multi-detector CT (MDCT) volumes. The simulator contains highly accurate, physics-based models of the full imaging chain and image formation process, comprising the following four main steps: (i) First, MDCT volumes were segmented to provide material-domain object models (e.g., soft-tissue, bone), which consist of voxel-wise fraction of each material. The object model was then used to estimate scatter and primary signal through MC simulation[1,2] and polyenergetic forward projection, respectively. These calculations included models of the system geometry, incident spectrum, focal spot blur, beam modulation, patient motion, antiscatter-grid, and detector response. (ii) An energy-dependent cascaded systems analysis model[10] was used to inject correlated quantum noise in the projection domain. The amount of injected noise was based on the number of photons and the incident spectrum (calculated based on the line integral for each material) at the surface of the detector for each pixel. (iii) Third, detector-domain artifacts including veiling glare and detector lag were added in the projection domain through spatial-temporal convolution with their associated kernel functions;[1,2] (iv)

Finally, realistic CBCT projection data were obtained by downsampling the projections to the specified readout binning mode and injecting uncorrelated electronic noise.

## 2.4 Experiments: CBCT of Low-Contrast Lesions

The synthesis network was trained using paired real MDCT and simulated CBCT images, which were the input and output of the simulator described in §2.3. Parameters for the simulation were adjusted to emulate the characteristics of a CBCT system common in image-guided surgery (the O2 O-arm, Medtronic; head scan protocol: 120 kV, 93 total mAs, 370 projections). A total of 22,000 slices were used for DL-Synthesis training (healthy and hydrocephalus subjects). Training was performed with the Adam optimizer (learning rate $5 \times 10^{-4}$, batch size = 8).

Two simulation studies involving low-contrast lesions (not present in the training set) were designed to investigate the performance of the proposed methods. Experiment #1 featured two types of simulated lesions added to a healthy patient: (i) simple circular lesions with varying contrast (-70 to +70 HU, pertinent to low-contrast features such as intracranial hemorrhage (ICH), ischemia, and abnormal fluid), size (diameter ranging from 10-40 mm), and location (random placement within the cranial vault); (ii) more complex star-polygon lesions with varying contrast [-70 to

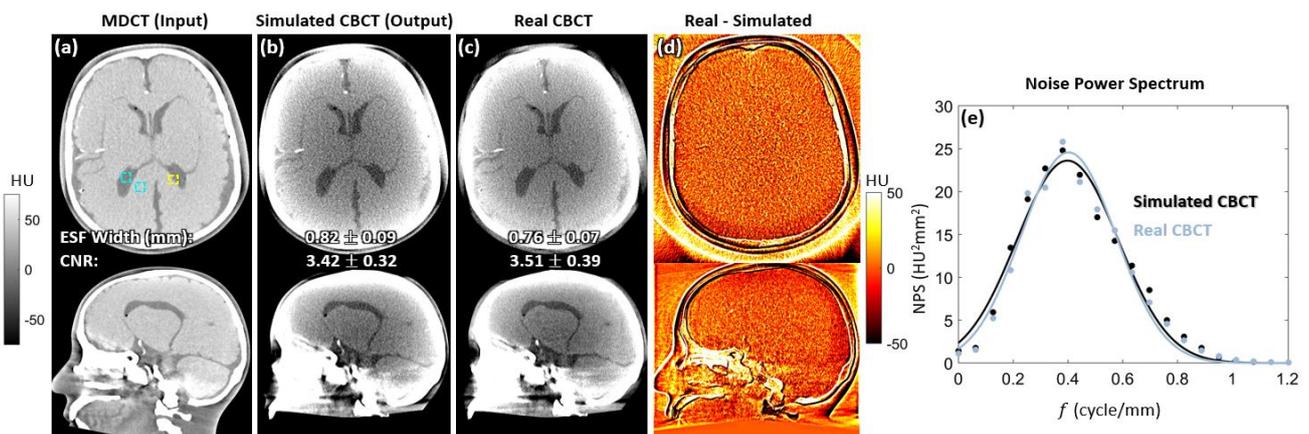

*Figure 4. Validation results for the high-fidelity forward simulator using an anthropomorphic head phantom. Top row: axial plane. Bottom row: sagittal plane. (a) MDCT images of the head phantom used as input to the simulator; (b) Simulated CBCT images. (c) Real CBCT images from the Medtronic O-Arm system. (d) Difference map. (e) Axial plane noise-power spectrum (NPS) for the simulated and real CBCT image. Resolution [edge-spread function (ESF) width] and contrast-to-noise ratio (CNR) for the simulated and real CBCT image were also labeled in (b) and (c).*

-20 HU and +20 HU to +70 HU], size (inner diameter ranging from 8 – 25 mm; outer diameter ranging from 20-60 mm), shape (number of vertices ranging from 3-12), and location (random within the brain parenchyma). Experiment #2 used a dataset featuring real hypodense lesions (~ −30 HU contrast) of edema and ischemia.

## 3 Results

### 3.1. Validation of Training Data Generation
The performance of the high-fidelity forward simulator is summarized in Fig. 4. Side-by-side comparison shows that the simulated CBCT accurately reproduces the measured experimental data acquired with the O-arm system. The high level of agreement is also illustrated by the difference map in (d). Quantitative measurement shows <10% discrepancy in spatial resolution, CNR, and NPS between the simulated and real CBCT images.

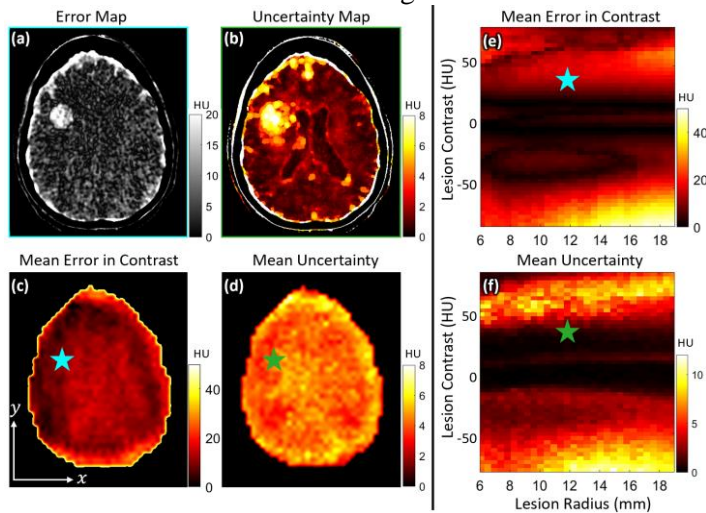

*Figure 5. Correlation between DL-Synthesis error and the statistical uncertainty in $\mu_s$ [Eq. (1)]. (a) Difference between the DL-Synthesis and ground truth for an example dataset in Exp #1 featuring a hyperdense lesion with a contrast of +40 HU. (b) Corresponding uncertainty map ( $\sigma$ ) for the soft-tissue region. (c) Mean error (in lesion contrast) computed as a function of lesion location. (d) Mean uncertainty within the lesion computed as a function of lesion location. Lesion contrast and radius were fixed to +40 HU and 12 mm, respectively, in (c-d). (e) Mean error computed as a function of lesion contrast and size. (f) Mean uncertainty within the lesion computed as a function of lesion contrast and size. Lesion location was the same as (a) for calculations in (e) and (f). Values at each point-pair in (c-d) or (e-f) show results from one dataset [e.g., The point-pair marked with blue-green stars corresponds to the dataset in (a-b)].*

### 3.2. Validation of Uncertainty Estimation
Figure 5 shows the uncertainty of DL-Synthesis for datasets with circular lesions in Exp #1. Taking as an example the dataset featuring a simulated hyperdense lesion (+40 HU contrast) as shown in Fig. 6(a), we see that uncertainty is highest in the region of the simulated lesion, since such a lesion was not present in the training set, leading to errors in the conventional DL-Synthesis prediction [Fig. 5(a)]. Note the correlation between error and uncertainty [Fig. 5(a-b)]. Additionally, the correlation of DL-Synthesis error and uncertainty is shown: (i) as a function of the location of the lesion in Fig. 5(c-d), and as a function of the size and contrast of the lesion in Fig. 5(e-f). We also observe clear correlation between synthesis error and uncertainty for lesions of different location, size, and contrast – evident in the similar pattern between (c-d) and between (e-f). For example, lesions adjacent to or within the lateral ventricles resulted in greater synthesis error and uncertainty as seen in (c-d). Uncertainty was therefore taken as a reasonable surrogate for regions susceptible to error in the synthesis image.

### 3.3. Experiment #1: Simulated Lesion
Figure 6 shows images reconstructed with conventional methods (FBP, PWLS, and DL-Synthesis) and the proposed DL-Recon methods with uncertainty information (DL-FBP and DL-MBIR) for an example dataset in Exp #1 (simulated hyperdense ICH lesion of +40 HU contrast). Compared to (artifact-corrected) FBP and PWLS, the DL methods show ~50% reduction in noise (at matched spatial resolution measured at the wall of the lateral ventricle) and ~53% improvement in image uniformity. The improved noise-resolution tradeoff of DL-MBIR is shown in Fig. 6(i). Unfortunately, the conventional DL-Synthesis method alone exhibits ~52% reduction in contrast of the ICH lesion (compared to truth), showing a lack of reliability / generalizability for structures unseen in the training data. The DL-FBP and DL-MBIR methods, on the other hand, are significantly more robust against such contrast reduction by weighting the physical measurements in regions of high uncertainty. DL-MBIR shows the expected advantages in noise-resolution tradeoffs compared to DL-FBP.

Figure 6(j) illustrates the importance of the spatially varying uncertainty penalty in DL-MBIR, where the performance of two variants is shown – one in which the penalty varies according to the uncertainty map and one in which $\beta$ is held constant [i.e., no uncertainty information, denoted as DL-MBIR (constant $\beta$ )]. With DL-Synthesis as a prior, the spatially varying, uncertainty-based model (DL-MBIR) maintains an accurate representation of lesion contrast via the data fidelity term. Incorporation of the physical model thus compensates for inaccuracies in DL-Synthesis in regions where uncertainty is high. Note that DL-MBIR with constant $\beta$ outperformed DL-Synthesis alone, showing the benefit of combining deep learning with physics models even when the uncertainty information is not available.

Figure 7 shows images reconstructed with conventional methods (FBP and DL-Synthesis) and the proposed DL-MBIR method for an example dataset in Exp #1 (hypodense star-polygon lesion with -40 HU contrast). Utilizing uncertainty information (highest in the hypodense lesion region), DL-Recon was able to mitigate biases introduced by the inaccuracy in DL-Synthesis (~45% improvement in lesion contrast and ~35% improvement in Dice coefficient) while maintaining the improved noise and uniformity characteristics of deep learning methods. The improvement in Dice coefficient for DL-MBIR reflects a higher degree of reliability in imaging pathologies unseen in the training data – an important aspect for many clinical scenarios.

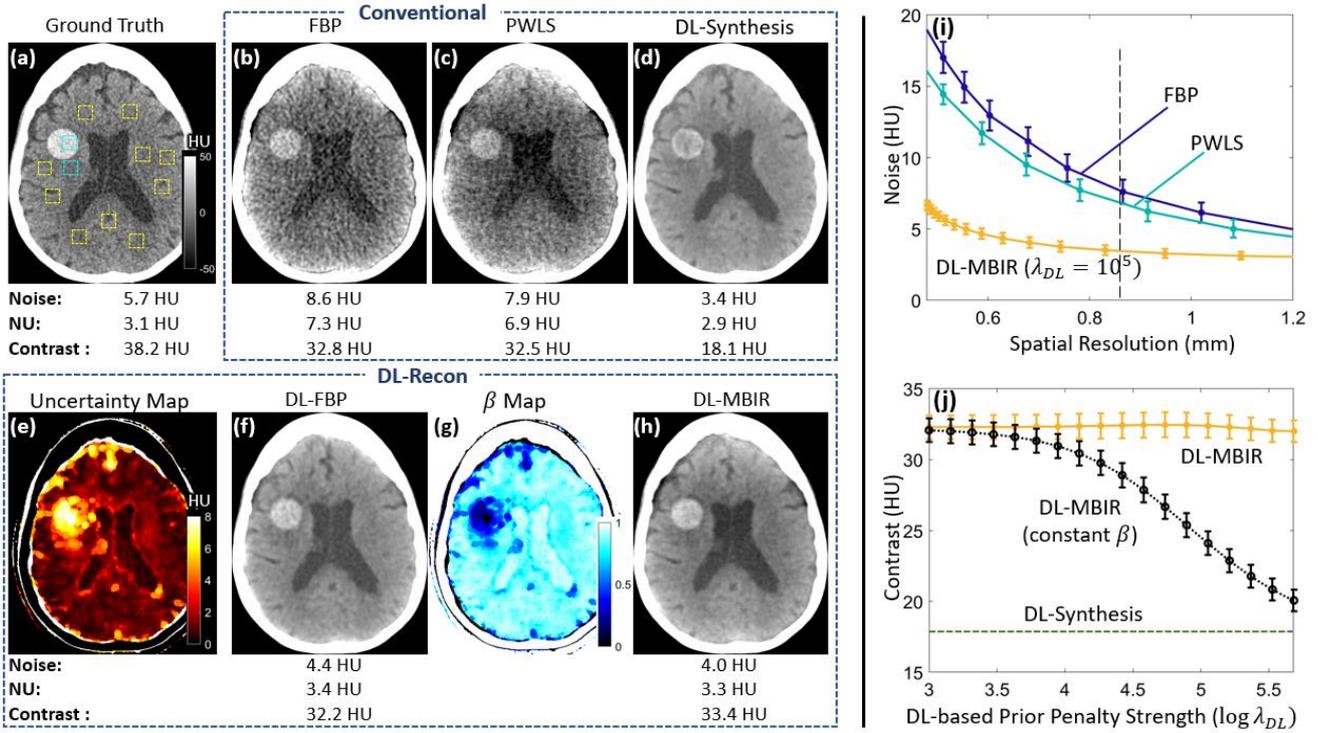

Figure 6. Image reconstructions and analysis for an example dataset in Exp #1. (a) Ground truth consisting of a clinical MDCT scan of the brain with the addition of a lesion of +40 HU contrast (ICH). (b) FBP. (c) PWLS. (d) DL-Synthesis. (e) Uncertainty map for DL-Synthesis. (f) DL-FBP. (g) Spatially varying penalty ("β map") computed by Eq. (3). (h) DL-MBIR. Note the more accurate ICH contrast in (f) [+32.2 HU] and (h) [+33.4 HU] compared to (d) [+18.1 HU] relative to truth (a) [+38.2 HU]. The spatial resolution in images (b, c, d, f, h) was matched at the boundary of the ventricle. The cyan ROIs in (a) were used to measure the lesion contrast, and the yellow ROIs were used to measure non-uniformity (NU) and noise in brain parenchyma. (i) Noise-resolution tradeoff for FBP, PWLS, and DL-MBIR. (j) ICH contrast for DL-MBIR with spatially varying beta and for DL-MBIR with constant β as a function of penalty strength $\lambda_{DL}$. Note that DL-MBIR maintains contrast despite the inaccurate DL-Synthesis prior.

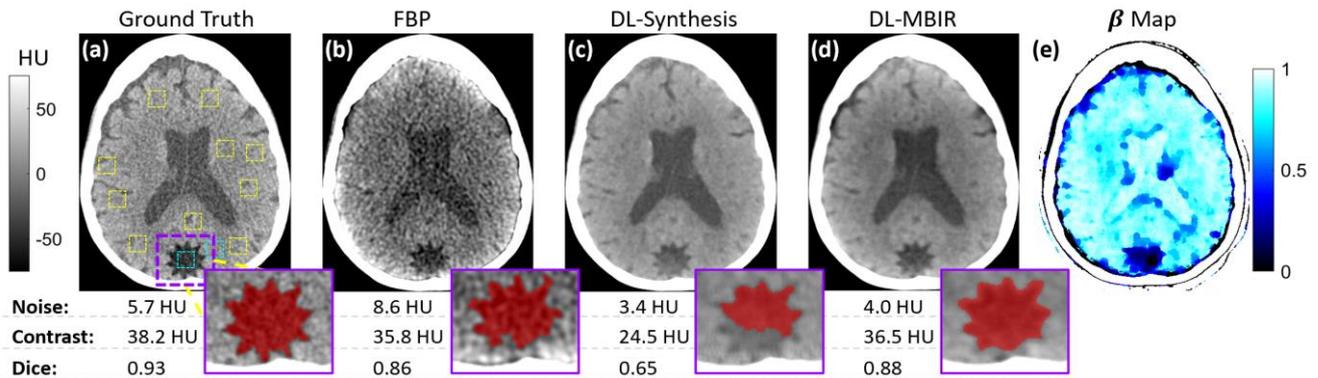

Figure 7. Reconstruction results for an example dataset in Exp #1, featuring a complex shaped (star-polygon) stimulus. (a) Ground truth consisting of a clinical MDCT scan of the brain with the addition of a hypodense lesion (ischemia) of -40HU contrast. (b) FBP. (c) DL-Synthesis. (d) DL-MBIR. (e) Spatially varying penalty ("β map") computed by Eq. (3). Dice coefficients from a threshold-based segmentation (threshold set to achieve the optimal segmentation in the FBP reconstruction) were measured within the purple ROI. Note the more accurate segmentation from DL-MBIR as compared to DL-Synthesis, allowing easier lesion analysis in a clinical workflow

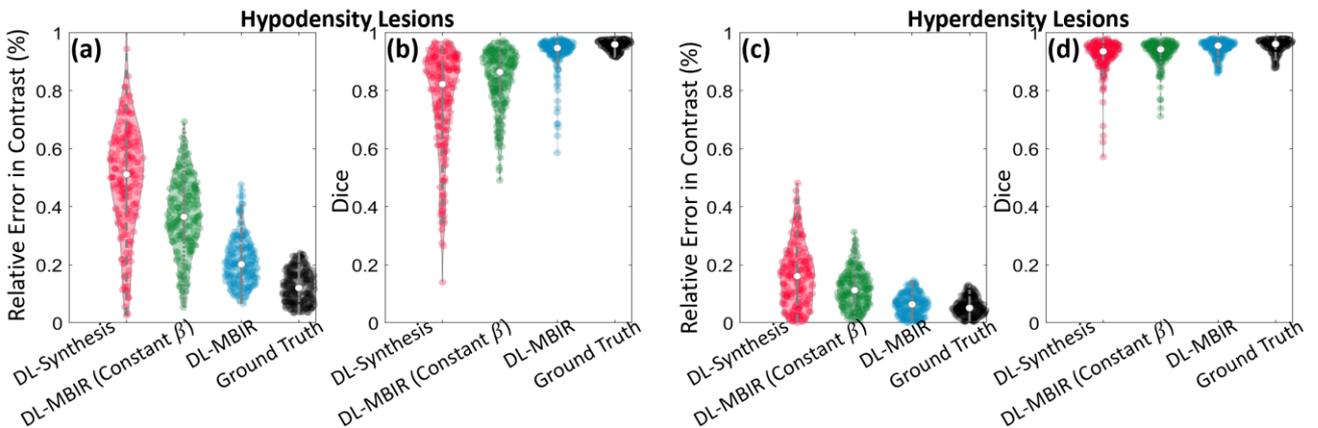

Figure 8. Quantitative analysis of reconstruction accuracy [DL-Synthesis, DL-MBIR (constant β), DL-MBIR, and ground truth] aggregated over all datasets in Exp #1 with star-polygon lesions. (a) Relative error in contrast for hyperdense lesions. (b) Dice coefficient for hyperdense lesions. (c) Relative error in contrast for hypodense lesions. (d) Dice coefficient for hypodense lesions.

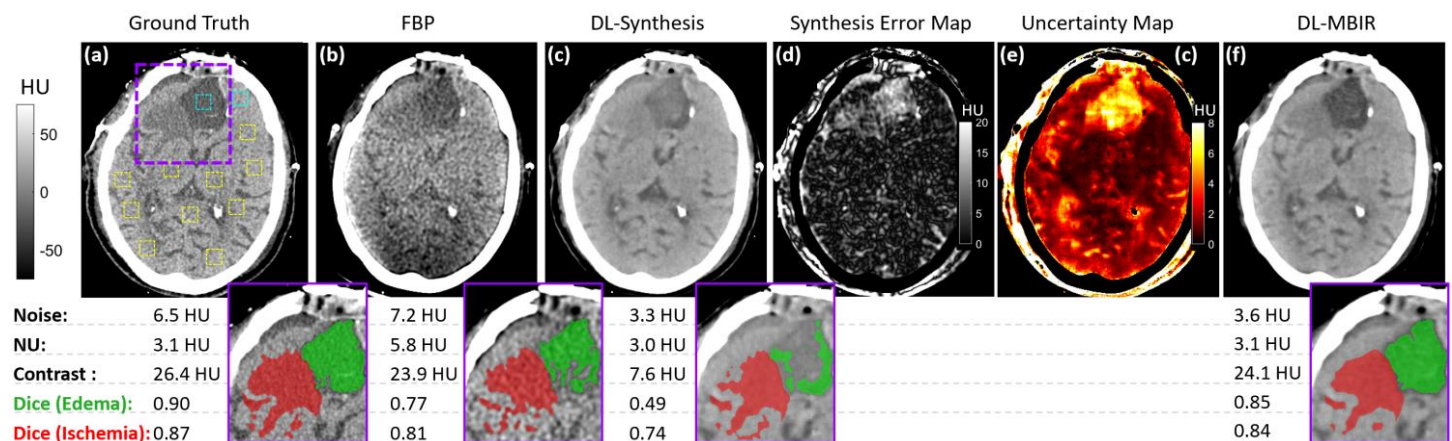

*Figure 9. Reconstruction results for Exp #2. (a) Ground truth consisting of a clinical MDCT scan of the brain with hypodense lesions. (b) FBP. (c) DL-Synthesis images. (d) Error map. (e) Uncertainty map for the DL-Synthesis image. (f) DL-MBIR. Note the improved lesion contrast in (e) and (f) compared to (c) [cyan ROIs]. The spatial resolution of (b, c, f) were matched at the boundary of the ventricle region (giving ESF width = 0.85 mm). Dice coefficients from a threshold-based segmentation were measured within the purple ROI for the edema (green) and ischemia lesion regions (red).*

Figure 8 shows the distribution of relative error in lesion contrast and Dice coefficient for all datasets in Exp #1 with star-polygon lesions. The results in Fig. 8 compare the performance of DL-Synthesis, DL-MBIR with spatially invariant (constant) $\beta$, and DL-MBIR methods. Compared with DL-Synthesis alone, DL-MBIR (constant $\beta$) improved the interquartile range (IQR) of the relative error in contrast and Dice coefficient for the hyperdense lesions by 31% and 10%, respectively, and DL-MBIR (spatially-varying beta) improved these characteristics by 51% and 23%, respectively. The performance of conventional DL-Synthesis was observed to be lower for reconstruction of hypodense lesions. Possible reasons for the decreased performance could lie in the similarity between hypodense lesions and most forms of CBCT artifacts, such as scatter and beam hardening, which tend to present as "hypodense" shading or streaks. With DL-MBIR, the IQR of the relative error in contrast and Dice coefficient for the hypodense lesions were improved by 58% and 68%, respectively, compared to DL-Synthesis.

## 3.4. Experiment #2: Real Pathology

Figure 9 shows images reconstructed with conventional methods (FBP and DL-Synthesis) and DL-MBIR. Compared to FBP, DL-MBIR shows ~50% reduction in noise and ~47% improvement in image uniformity throughout the brain parenchyma. The DL-Synthesis network exhibited highest uncertainty in the region of hypodense lesions, leading to inaccurate representation of the edematous lesion (~42% reduction in Dice; green overlay) and the ischemic lesion (~12% reduction in Dice; red overlay). By including uncertainty information and physics-based reconstruction models, DL-MBIR accurately depicted the contrast and shape of the lesions while maintaining the improved noise and uniformity characteristics of DL-Synthesis.

## 4 Discussion & Conclusion

This work presented a new type of DL-based image reconstruction method (termed DL-Recon) that integrates physics-based models with image synthesis based on epistemic uncertainty. To our knowledge, this represents a novel incorporation of Bayesian uncertainty in a neural network approach with physics-based and DL-based CBCT image reconstruction. Two variations of DL-Recon were proposed in this work, both maintaining the basic advantages of conventional DL-Synthesis: (i) the DL-FBP method improved the accuracy of reconstruction and offers practical advantages of runtime efficiency; and (ii) the DL-MBIR offered further image quality improvement due to the more accurate physical model and the explicit data-fidelity constraint. Compared with DL-Synthesis alone, both of the DL-Recon methods showed improved robustness to anatomical variations (e.g., pathologies) that were unseen in the training set. Besides image reconstruction, other applications of synthehsis uncertainty can be envisioned, for example: helping to identify abnormal anatomy and image features in image classifcaiton; providing a quantitative measurement of sufficiency in the size and/or variety of a training dataset; and helping to quantify improvements (or unexpected variations) in continuous learning. Ongoing work includes extension to fully 3D image reconstruction and investigation in clinical studies.

## References

[1] Sisniega, A., et al. "Image quality, scatter, and dose in compact CBCT systems with flat and curved detectors." Medical Imaging 2018: Physics of Medical Imaging. Vol. 10573. International Society for Optics and Photonics, 2018.

[2] Wu P et al. Cone-beam CT for imaging of the head/brain: Development and assessment of scanner prototype and reconstruction algorithms. Med Phys. 2020;47(6):2392-2407. doi:10.1002/mp.14124

[3] Chen H et al. Low-Dose CT with a Residual Encoder-Decoder Convolutional Neural Network. IEEE TMI. 2017;36(12):2524-2535.

[4] Yang Q et al. Low-Dose CT Image Denoising Using a Generative adversarial Network with Wasserstein Distance and Perceptual Loss. IEEE TMI. 2018;37(6):1348-1357.

[5] Wu D et al. Iterative low-dose CT reconstruction with priors trained by artificial neural network. IEEE TMI. 2017;36(12):2479-2486.

[6] Zhang C et al. Deep learning enabled prior image constrained compressed sensing reconstruction framework. SPIE Medical Imaging 2020 Vol 11312.

[7] Gal Y et al. Dropout as a Bayesian approximation: Representing model uncertainty in deep learning. ICML 2016:1050-1059.

[8] Isola P et al. Image-to-image translation with conditional adversarial networks. CVPR 2017 1125–1134.

[9] Fessler JA. Penalized weighted least-squares image reconstruction for positron emission tomography. IEEE TMI. 1994;13(2):290-300.

[10] Siewerdsen JH et al. Empirical and theoretical investigation of the noise performance of indirect detection, active matrix flat-panel imagers for diagnostic radiology. Med Phys. 1997;24(1):71-89